\documentclass[a4paper,12pt]{article}  
\usepackage{latexsym}  
\usepackage{amsfonts} 
\usepackage{multirow} 
\usepackage{array} 
\usepackage{tabularx}  
\usepackage{amsmath}  
\usepackage[dvips]{graphicx}  
\begin{document}  
\pagestyle{empty}  
\addtolength{\voffset}{-2cm}  
\setlength{\textheight}{640pt}  
\begin{flushright}  
hep-th/9808125 \\  
TUW-98-18  
\end{flushright}  
\vspace{3 cm}  
\begin{center}  
\LARGE{Duality for $SU \! \times \! SO$ and $SU \! \times \! Sp$ via  
Branes} \\  
\vspace{1.8 cm}  
\large{Esperanza Lopez
and Barbara Ormsby} \\  
\vspace{0.5 cm}  
\normalsize{Institut f\"ur theoretische Physik, TU-Wien \\  
Wiedner Hauptstra\ss e 8-10 \\  
A-1040 Wien, Austria \\  
\texttt{elopez@tph16.tuwien.ac.at \\  
bhackl@tph16.tuwien.ac.at}} \\  
\end{center}  
\vspace{2cm}  
Using a six-orientifold, fourbranes and four fivebranes in type IIA 
string theory we construct $\mathcal{N}$=1  
supersymmetric gauge theories in four dimensions with product group 
$SU(M)\times SO(N)$ or $SU(M)\times Sp(2N)$, a bifundamental flavor and  
quarks. We obtain the Seiberg dual for these theories and rederive it 
via branes. To obtain the complete dual group via branes we have to 
add semi-infinite fourbranes. We propose that the theory 
derived from branes has a meson deformation switched on. This 
deformation implies higgsing in the dual theory. The addition of 
the semi-infinite fourbranes compensates this effect.

\newpage  
\pagestyle{plain}  
\newpage  
\section{Introduction}  
  
Dirichlet branes have proved to be an extremely useful tool for   
obtaining non-perturbative information about gauge theories.   
A variety of brane constructions are available, which allow to  
induce a wide spectrum of gauge theories in the world-volume of  
the Dirichlet branes. We will be interested in configurations of  
Dirichlet fourbranes ending on Neveu-Schwarz fivebranes in Type   
IIA string theory. These configurations can be organized such that  
some supersymmetry is preserved \cite{hw}. When all the fivebranes  
are parallel, $1/4$ of the initial supersymmetries survives   
and the theory describing the low energy effective action on the  
branes is an $\mathcal{N}$=2 four-dimensional gauge theory.  
One can reduce further to $\mathcal{N}$=1 by rotating the fivebranes  
in the orthogonal directions an $SU(2)$ angle \cite{rotbrane}.   
Quantum corrections can be incorporated in this picture by lifting  
the brane configuration to M-theory \cite{wM}. Then the singular   
intersections  
of fivebranes and fourbranes are smoothed out and we obtain a single  
M-fivebrane wrapped around a Riemann surface with four   
uncompactified world-volume dimensions. For configurations preserving  
1/4 of the supersymmetry this Riemann surface coincides with the   
Seiberg-Witten curve describing the Coulomb branch of $\mathcal{N}$=2  
gauge theories \cite{sw}. Even in the Type IIA framework, the brane   
construction  
of gauge theories allows to derive very non-trivial information.   
The Seiberg dual of a given $\mathcal{N}$=1 gauge theory \cite{sei} can   
be obtained from certain brane moves \cite{elgvkt}.  
  
We will derive the Seiberg dual for an $\mathcal{N}$=1  
$SU(M) \times SO(N)$ or $SU(M) \times Sp(2N)$ gauge theory with a   
bifundamental flavor and quarks. This case has not yet been analyzed.  
We will obtain the dual theory first by field theory considerations   
and then by brane moves.  
This case offers a nice check for the brane approach to gauge theories.   
In order to obtain many of the known dual pairs, a superpotential 
must be added to the electric theory which truncates the set of 
chiral operators. On the other hand, the only possible obstruction to
perform the mentioned brane moves is when we have to cross parallel
branes.
We find in our case a one to one correspondence between configurations with 
several parallel branes and situations in which the superpotential does not 
truncate the chiral ring. This will be the subject of sections 2 and 3.  

In \cite{hb} the Seiberg dual for an $SU(N_1) \times SU(N_2) \times  
SU(N_3)$ was rederived via branes. They observed that branes  
predicted a smaller dual group. However the mismatch could be cured by  
adding a number of fourbranes to the dual configuration without  
modifying the linking numbers. We will encounter a similar problem.   
We will propose that the addition of fourbranes can be understood  
as a reverse of higgsing. In section 4 we will find a deformation of the  
electric theory which higgses the dual magnetic theory to the result  
derived from branes. We justify why such a deformation should be switched 
on by analyzing brane moves which correspond to dualize a single
factor group.
We treat in detail the case $SU(M) \times SO(N)$ and in section 5 extend 
briefly the results to $SU(M) \times Sp(2N)$.  
  
\section{Brane Configuration} \label{bc}  
  
\begin{figure}[hb]  
\begin{center}  
\includegraphics[angle=0, width=0.8\textwidth]{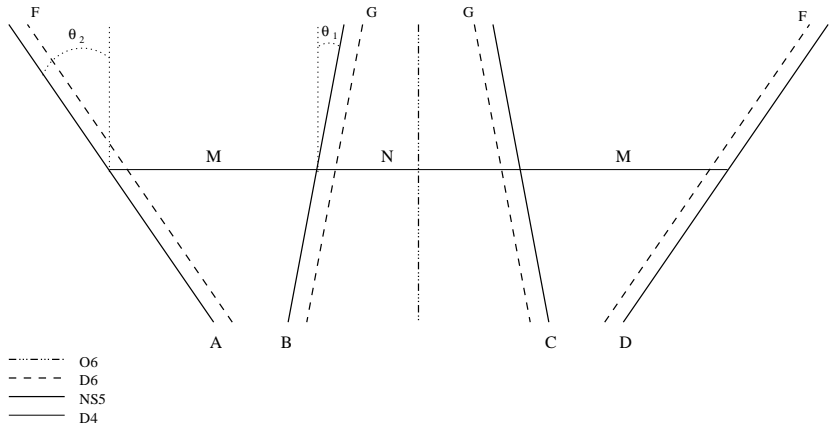}  
\caption{The brane configuration corresponding to an $\mathcal{N}$=1 $SU(M)  
\times SO(N)$ theory.}  
\label{nauf1}  
\end{center}  
\end{figure}

\noindent Our first ingredient is an orientifold sixplane extending in the   
$(x_0,x_1,x_2,x_3,$ $x_7,x_8,x_9)$ directions. Thus the brane  
configurations  
which we will consider must be $\mathbb{Z}_2$ symmetric in the 
$(x_4,x_5,x_6)$  
directions. We will use four NS-fivebranes with world-volume along  
$(x_0,x_1,x_2,x_3,x_4,x_5)$ and fourbranes suspended between them  
expanding in $(x_0,x_1,x_2,x_3)$ and with finite extent in the $x_6$ direction.   
In order to obtain gauge theories with $\mathcal{N}$=1 supersymmetry in   
the macroscopic dimensions of the fourbranes, we rotate the fivebranes  
an $SU(2)$ angle from $(x_4,x_5)$ towards $(x_8,x_9)$ \cite{rotbrane}.  
The leftmost fivebrane A will be tilted at an angle $\theta_2$, the 
interior fivebrane B at an angle $\theta_1$. We place $M$   
fourbranes between the A and B fivebranes and $N$ fourbranes between B and   
its mirror C. In addition there will be $F$ sixbranes parallel to  
the A fivebrane and $G$ sixbranes parallel to the B fivebrane in the   
$(x_4,x_5,x_8,x_9)$ space and extending in $(x_0,x_1,x_2,x_3,x_7)$.   
The rest of the configuration is determined by the $\mathbb{Z}_2$ 
action of the  
orientifold (see Fig.\ref{nauf1}).   
For an orientifold sixplane of positive Ramond charge, this configuration  
gives rise to an $\mathcal{N}$=1 gauge theory with product group   
$SU(M)\times SO(N)$ and the following matter content: fields $X$ and   
$\widetilde X$ forming a flavor in the bifundamental representation,   
$F$ flavors $Q$, $\widetilde Q$ transforming in the fundamental 
representation   
of $SU(M)$ and $2G$ chiral fields $Q'$ in the   
vector representation of $SO(N)$ \cite{ll}. For an orientifold sixplane of   
negative Ramond charge we obtain a gauge theory with group 
$SU(M)\times Sp(2N)$  and the same matter content.  
We will concentrate on the product group $SU(M)\times SO(N)$. The analysis  
of $SU(M)\times Sp(2N)$ will be very similar and we will just make some  
remarks referring to it in section 5.   

We want to determine the superpotential associated to our gauge theory.  
A convenient approach is to take as reference the superpotential for a  
brane configuration with additional massless tensor matter, add the mass 
term for the tensor field implied by the fivebrane rotation and  
integrate out this field. We will do the analysis separately for the 
tensor fields coming from the $SU(M)$ and $SO(N)$ sectors.  
The fourbranes suspended between the A and B fivebranes give rise to   
the $SU(M)$ factor group. When $\theta_1=\theta_2$ there is an 
additional massless $SU(M)$ adjoint field, $\phi$, whose expectation values 
move along the world-volume of the fivebranes. 
For arbitrary angles the superpotential for the $SU$ sector will  
be $W= X \phi {\widetilde X} + \mu \mbox{Tr}\phi^2$, where the field
$\phi$   
gets a mass $\mu=\mbox{tan} (\theta_2 - \theta_1)$ \cite{rotbrane}. There 
is no coupling   
between $\phi$ and the $F$ quarks since we are considering sixbranes  
parallel to the A fivebrane. Integrating out $\phi$ we get a superpotential   
\begin{equation}  
W_{SU} = \frac{-1}{4 \mbox{tan} (\theta_2 - \theta_1) } \left( \mbox{Tr}  
( X {\widetilde X} )^2 - \frac{1}{M} (\mbox{Tr} X {\widetilde X})^2 \right).  
\end{equation}    
  
Fourbranes between the B and the C fivebrane give rise to the  
$SO(N)$ factor group. When $\theta_1=0$ we   
have an additional massless chiral field in   
the adjoint representation, $\phi_A$. When $\theta_1= \pi/2$ the B fivebrane  
and its dual are also parallel and we get additional massless matter,  
transforming this time in the symmetric representation of $SO(N)$, 
$\phi_S$ \cite{witold}.    
The bifundamental field couples to both $\phi_A$ and $\phi_S$, which for  
arbitrary $\theta_1$ are massive. The associated   
superpotential for the $SO$ sector is  
\begin{equation}  
W = X \phi_A {\widetilde X} + X \phi_S {\widetilde X} + \mu' 
\mbox{Tr}\phi_A^2 -   
\frac{1}{\mu'} \mbox{Tr} \phi_S^2,   
\end{equation}   
where $\mu'=\mbox{tan} \theta_1$ \footnote{The mass for $\phi_S$ is   
$\mbox{tan} (\pi/2 + \theta_1)=-1/{\mu'}$.}. Since the $2G$ $SO(N)$ chiral 
vector fields come from sixbranes parallel to the fivebranes, we will
again suppose that there is no coupling between them and $\phi_A$, 
$\phi_S$. We will present a more careful discussion of
this point in section 4. Integrating out both tensor fields we get  
\begin{equation}  
W_{SO} = \frac{-1}{4 \mbox{tan} 2 \theta_1} \mbox{Tr}( X {\widetilde X} )^2  
+ \frac{1}{4 \mbox{sin} 2 \theta_1} \mbox{Tr} X {\widetilde X} 
{\widetilde X} X.  
\end{equation}   
The final answer for the superpotential is then   
\begin{equation}  
W= W_{SU} + W_{SO} = a \mbox{Tr}( X {\widetilde X} )^2 + b  
\mbox{Tr} X {\widetilde X} {\widetilde X} X + c (\mbox{Tr} X 
{\widetilde X} )^2,  
\label{w}  
\end{equation}   
where  
\begin{equation}  
a=-\frac{1}{4} \! \left( \! \frac{1}{\mbox{tan} (\theta_2 \! - \! \theta_1)}  
+ \frac{1}{\mbox{tan} 2 \theta_1} \! \right),  
\hspace{3mm} b=\frac{1}{4 \mbox{sin} 2 \theta_1}, \hspace{3mm}  
c=\frac{1}{4 M \mbox{tan} (\theta_2 \! - \! \theta_1)}.  
\label{values}  
\end{equation}  
  
Using the F-term equations for the superpotential  
\begin{eqnarray}  
a\widetilde{X}X\widetilde{X}+b\widetilde{X}\widetilde{X}X+  
c\widetilde{X} \: \mbox{Tr} X\widetilde{X}  &=&0, \label{f1} \\  
aX\widetilde{X}X+bXX\widetilde{X}+c X \: \mbox{Tr} X\widetilde{X}&=&0, 
\nonumber  
\end{eqnarray}  
we can deduce the chiral mesons of the theory. For generic values of the  
coefficients $a$ and $b$ we get:  
$M_0=Q \widetilde{Q}$, $M_1=Q \widetilde{X} X \widetilde{Q}$,   
$M_2=Q \widetilde{X} X \widetilde{X} X \widetilde{Q}$,    
$M'_0=Q' Q'$, $M'_1=Q' X \widetilde{X} Q' $,   
$M'_2=Q' X \widetilde{X} \widetilde{X} X Q'$,   
$P_0=Q \widetilde{X} Q'$,   
$P_1=Q \widetilde{X} X \widetilde{X} Q'$,  
$\widetilde{P}_0=\widetilde{Q} X Q'$,   
$\widetilde{P}_1=\widetilde{Q} X \widetilde{X} X Q$,   
$R_1=Q \widetilde{X} \widetilde{X} Q$,  
$\widetilde{R}_1=\widetilde{Q} X X \widetilde{Q}$.   
For particular values of $a$ and $b$ the F-term equations can fail   
to truncate the chiral mesons to a finite set. This situation occurs when   
$a=0$; then mesons containing $(\widetilde{X} X)^k$ are allowed for any $k$.  
Using (\ref{values}) this corresponds to $\theta_1=- \theta_2$, i.e. when  
the A and C fivebranes are parallel.  
In order to discard mesons containing the combination  
$\widetilde{X} X {\widetilde X} \widetilde{X}$ it was necessary to use 
\begin{equation} \label{rel}  
(a^2 - b^2) \widetilde{X}X {\widetilde X}  
\widetilde{X} = c (b-a) {\widetilde X} \widetilde{X}\: \mbox{Tr}  
X\widetilde{X},   
\end{equation}  
which can be easily deduced from the F-term equations.    
Analogous relations hold for $\widetilde{X} {\widetilde X} X \widetilde{X}$,
$X {\widetilde X} XX$ and $XX {\widetilde X} X$ .
These relations are trivial identities when $a= b$. When $a=-b$ they just  
imply that the product of some mesons with $\mbox{Tr} X  
{\widetilde X}$ is not a chiral primary. From (\ref{values}), $a=b$  
implies $\theta_2=0$ and $a=-b$ implies $\theta_2=\pi/2$. These two  
situations correspond to the leftmost fivebrane A and its mirror D being  
parallel.  
When the A and B fivebranes are parallel, or B and C are parallel, 
we get additional tensor fields becoming massless and the set of chiral 
mesons also changes. Thus the stated set of chiral mesons is only valid 
when there are no parallel fivebranes.   
 
In the next sections we will be interested in obtaining the Seiberg  
dual for our $\mathcal{N}$=1 $SU(M) \times SO(N)$   
theory. We will derive it first from field theory methods and then from  
brane moves. The field theory derivation will be valid only when the  
set of chiral mesons is that stated above. On the other hand, the brane moves  
necessary to recover the dual theory will involve reversing the order of all
fivebranes and sixbranes and this process is only well defined when there 
are no parallel fivebranes in our configuration. Notice that in our
set-up parallel fivebranes imply sixbranes parallel to more than one 
fivebrane.
Since the derived finite set of chiral 
mesons is valid if and only if there are no parallel fivebranes, we have  
an additional check for the validity of the brane derivation of Seiberg  
dualities.    
 
The anomaly-free global symmetry group of our theory is  
\begin{equation} \label{globsym}  
SU(F)_L\times SU(F)_R\times SU(2G)\times U(1)_R\times U(1)_B\times U(1)_X    
\end{equation}   
The brane diagram of Fig.\ref{nauf1} does not exhibit the full $SU(2G)$  
flavor symmetry.
By bringing sixbranes over fivebranes we could obtain at  
most $SU(G) \times SU(G)$ \cite{hb}. However (\ref{globsym}) is the global 
symmetry group for the superpotential (\ref{w}) and thus it is the group  
we should consider in deriving the Seiberg dual theory. Another restriction 
of the brane construction is that we always get an even number of $SO$  
vectors, $2G$. The results of next section are valid for the general case  
where the $SO$ factor has $G'$ vectors just by substituting $2G$ by  
$G'$.  
The transformation properties of the matter fields under the gauge and  
global symmetry groups are summarized in table \ref{t2}.   
 
\setlength{\extrarowheight}{5pt}   
\begin{table}[hbt] 
\begin{small}   
\begin{center}   
\setlength{\tabcolsep}{1mm}  
\begin{tabular}{|c||c|c|c|c|c|c|c|c|}     
\hline    
 & SU($M$) & SO($N$) & SU($F$)$_L$ & SU($F$)$_R$ & SU($2G$) & U(1)$_R$ & 
U(1)$_B$ & U(1)$_X$ \\   
\hline \hline   
$Q$ & $M$ & 1 & $F$ & 1 & 1 & $1-\frac{2M-N}{2F}$ & $\frac{1}{M}$ & 0  \\   
\hline   
$\widetilde{Q}$ & $\overline{M}$ & 1 & 1 & $F$ & 1 & $1-\frac{2M-N}{2F}$ & 
-$\frac{1}{M}$ & 0 \\   
\hline   
$Q'$ & 1 & $N$ & 1 & 1 & $2G$ & $1-\frac{N-M-2}{2G}$ & 0 & 0 \\   
\hline   
$X$ & $M$ & $N$ & 1 & 1 & 1 & $\frac{1}{2}$ & $\frac{1}{M}$ & 1 \\   
\hline   
$\widetilde{X}$ & $\overline{M}$ & $N$ & 1 & 1 & 1 & $\frac{1}{2}$ & 
-$\frac{1}{M}$ & -1 \\   
\hline   
\end{tabular}   
\caption{Matter content of the electric theory.} \label{t2}    
\end{center} 
\end{small}   
\end{table}

\section{Dual Theory}  
  
We propose that the dual theory has gauge group $SU({\widetilde M}) \times   
SO({\widetilde N})$ with $\widetilde{M}=4F+4G-M+4$ and 
$\widetilde{N}=8G+4F-N+8$.  
The matter content is given by fields $Y$ and $\widetilde Y$ forming a   
flavor in the bifundamental representation, $F$ flavors $q$, $\widetilde q$   
transforming in the fundamental representation of $SU({\widetilde M})$ and   
$2G$ fields $q'$ in the vector representation of   
$SO({\widetilde N})$. In addition there will be singlets $M_i,M'_i$ with   
$i=0,1,2$, $P_j,\widetilde{P_j}$ with $j=0,1$  
and $R_1,\widetilde{R}_1$ in one to one correspondence with the chiral   
mesons of the electric theory. The matter fields transform under the   
symmetries as indicated in table \ref{t3}.  
  
The dual superpotential is   
  \begin{eqnarray}  \label{spdual}  
W&=& \mbox{Tr} (Y\widetilde{Y})^2+\mbox{Tr} Y\widetilde{Y}\widetilde{Y}Y  
+(\mbox{Tr}Y\widetilde{Y})^2+M_0\widetilde{q}(Y\widetilde{Y})^2q+
M_1\widetilde{q}Y\widetilde{Y}q \nonumber \\   
&+&M_2\widetilde{q}q+M'_0q'Y\widetilde{Y}\widetilde{Y}Yq'+M'_1q'Y
\widetilde{Y}q'+M'_2q'q'+P_0q\widetilde{Y}Y\widetilde{Y}q'\\    
&+&P_1q\widetilde{Y}q'+\widetilde{P_0}\widetilde{q}Y\widetilde{Y}Yq'+
\widetilde{P_1}\widetilde{q}Yq'+R_1q\widetilde{Y}\widetilde{Y}q+
\widetilde{R_1}\widetilde{q}YY\widetilde{q}, \nonumber  
  \end{eqnarray}   
where for simplicity we have ignored (dimensionfull) coefficients in front 
of each term.  
 
\begin{table}[hbt!]   
\begin{small} 
\begin{center}   
\setlength{\tabcolsep}{1mm}  
\begin{tabular}{|c||c|c|c|c|c|c|c|c|}     
\hline    
 & SU(${\widetilde M}$) & SO(${\widetilde N}$) & SU($F$)$_L$ & SU($F$)$_R$ 
& SU($2G$) & U(1)$_R$ & U(1)$_B$ & U(1)$_X$ \\   
\hline \hline   
$q$ & $\widetilde{M}$ & 1 & $\overline{F}$ & 1 & 1 & 
$1-\frac{2\widetilde{M}-\widetilde{N}}{2F}$ & 
$\frac{1}{\widetilde{M}}$ &$\frac{\widetilde{M}+M-2F}{\widetilde{M}}$  \\   
\hline   
$\widetilde{q}$ & $\overline{\widetilde{M}}$ & 1 & 1 & $\overline{F}$ & 
1 & $1-\frac{2\widetilde{M}-\widetilde{N}}{2F}$ & 
-$\frac{1}{\widetilde{M}}$ & $-\frac{\widetilde{M}+M-2F}{\widetilde{M}}$  \\   
\hline   
$q'$ & 1 & $\widetilde{N}$ & 1 & 1 & $\overline{2G}$ & 
$1-\frac{\widetilde{N}-\widetilde{M}-2}{2G}$ & 0 & 0 \\   
\hline   
$M_0$ & 1 & 1 & $F$ & $F$ & 1 & $-2+\frac{2\widetilde{M}-
\widetilde{N}}{F}$ & 0 & 0 \\   
\hline   
$M_1$ & 1 & 1 & $F$ & $F$ & 1 & $-1+\frac{2\widetilde{M}-
\widetilde{N}}{F}$ & 0 & 0 \\   
\hline   
$M_2$ & 1 & 1 & $F$ & $F$ & 1 & $\frac{2\widetilde{M}-
\widetilde{N}}{F}$ & 0 & 0 \\   
\hline   
$M'_0$ & 1 & 1 & 1 & 1 & sym & $-2+\frac{\widetilde{N}-
\widetilde{M}-2}{G}$ & 0 & 0\\   
\hline  
\multirow{2}{1cm}{\hspace{0.3cm}$M'_1$} & 
\multirow{2}{1cm}{\hspace{0.3cm} 1} & 
\multirow{2}{1cm}{\hspace{0.3cm} 1} & 
\multirow{2}{1cm}{\hspace{0.3cm} 1} & 
\multirow{2}{1cm}{\hspace{0.3cm} 1} & sym $\! \oplus$ & 
\multirow{2}{2.5cm}{\hspace{1mm}$-1+\frac{\widetilde{N}-
\widetilde{M}-2}{G}$} & \multirow{2}{1cm}{\hspace{0.4cm}0} & 
\multirow{2}{1cm}{\hspace{0.4cm}0}\\    
 & & & & & asym & & & \\   
\hline   
$M'_2$ & 1 & 1 & 1 & 1 & sym & $\frac{\widetilde{N}-\widetilde{M}-2}{G}$ 
& 0 & 0 \\   
\hline   
\multirow{2}{1cm}{\hspace{0.3cm}$P_0$} & 
\multirow{2}{1cm}{\hspace{0.3cm} 1} & 
\multirow{2}{1cm}{\hspace{0.3cm} 1} & 
\multirow{2}{1cm}{\hspace{0.3cm} $F$} & 
\multirow{2}{1cm}{\hspace{0.3cm} 1} & 
\multirow{2}{1cm}{\hspace{0.2cm} $2G$} & 
$-\frac{3}{2}+\frac{2\widetilde{M}-\widetilde{N}}{2F}+$ 
& \multirow{2}{1cm}{\hspace{0.3cm} 0} & 
\multirow{2}{1cm}{\hspace{0.15cm} -1}\\    
 & & & & & & $\frac{\widetilde{N}-\widetilde{M}-2}{2G}$ & &   \\   
\hline   
\multirow{2}{1cm}{\hspace{0.3cm}$P_1$} & 
\multirow{2}{1cm}{\hspace{0.3cm} 1} & 
\multirow{2}{1cm}{\hspace{0.3cm} 1} & 
\multirow{2}{1cm}{\hspace{0.3cm} $F$} & 
\multirow{2}{1cm}{\hspace{0.3cm} 1} & 
\multirow{2}{1cm}{\hspace{0.2cm} $2G$} & 
$-\frac{1}{2}+\frac{2\widetilde{M}-\widetilde{N}}{2F}+$ & 
\multirow{2}{1cm}{\hspace{0.3cm} 0} & 
\multirow{2}{1cm}{\hspace{0.15cm} -1} \\   
 & & & & & & $\frac{\widetilde{N}-\widetilde{M}-2}{2G}$ & & \\   
\hline   
\multirow{2}{1cm}{\hspace{0.3cm}$\widetilde{P}_0$} & 
\multirow{2}{1cm}{\hspace{0.3cm} 1} & 
\multirow{2}{1cm}{\hspace{0.3cm} 1} & 
\multirow{2}{1cm}{\hspace{0.3cm} 1} & 
\multirow{2}{1cm}{\hspace{0.3cm} $F$} & 
\multirow{2}{1cm}{\hspace{0.3cm} $2G$} & 
$-\frac{3}{2}+\frac{2\widetilde{M}-\widetilde{N}}{2F}+$ & 
\multirow{2}{1cm}{\hspace{0.3cm} 0} & \multirow{2}{1cm}{\hspace{0.3cm} 1} \\   
 & & & & & & $\frac{\widetilde{N}-\widetilde{M}-2}{2G}$ & & \\   
\hline   
\multirow{2}{1cm}{\hspace{0.3cm}$\widetilde{P}_1$} & 
\multirow{2}{1cm}{\hspace{0.3cm} 1} & 
\multirow{2}{1cm}{\hspace{0.3cm} 1} & 
\multirow{2}{1cm}{\hspace{0.3cm} 1} & 
\multirow{2}{1cm}{\hspace{0.3cm} $F$} & 
\multirow{2}{1cm}{\hspace{0.3cm} $2G$} & 
$-\frac{1}{2}+\frac{2\widetilde{M}-\widetilde{N}}{2F}+$ & 
\multirow{2}{1cm}{\hspace{0.3cm} 0} & \multirow{2}{1cm}{\hspace{0.3cm} 1} \\   
 & & & & & & $\frac{\widetilde{N}-\widetilde{M}-2}{2G}$ & & \\   
\hline   
$R_1$ & 1 & 1 & sym & 1 & 1 & $-1+\frac{2\widetilde{M}-\widetilde{N}}{F}$ 
& 0 & -2 \\   
\hline   
$\widetilde{R}_1$ & 1 & 1 & 1 & sym & 1 & $-1+\frac{2\widetilde{M}-
\widetilde{N}}{F}$ & 0 & 2 \\   
\hline   
$Y$ & $\widetilde{M}$ & $\widetilde{N}$ & 1 & 1 & 1 & $\frac{1}{2}$ & 
$\frac{1}{\widetilde{M}}$ & $\frac{M-2F}{\widetilde{M}}$ \\   
\hline 
$\widetilde{Y}$ & $\overline{\widetilde{M}}$ & $\widetilde{N}$ & 1 & 1 & 
1 & $\frac{1}{2}$ & -$\frac{1}{\widetilde{M}}$ & 
$-\frac{M-2F}{\widetilde{M}}$ \\   
\hline   
\end{tabular}   
\caption{Matter content of the magnetic theory.} \label{t3}    
\end{center}   
\end{small} 
\end{table}   
   
The usual test for the duality ansatz are the t'Hooft anomaly matching   
conditions for the global symmetry group. The anomalies computed   
with the fermions of the electric theory must match those computed with the   
fermions of the magnetic theory. Indeed we find for both theories    

\vspace{3mm}  
\begin{tabular}{r l}   
U(1)$_R$ & $-M^2+MN-\frac{N^2}{2}+\frac{3N}{2}-1$ \\   
U(1)$_R^3$ & $-2MF(\frac{2M-N}{2F})^3-2GN(\frac{N-M-2}{2G})^3-
\frac{1}{4}MN$\\   
 & $+M^2-1+\frac{N(N-1)}{2}$ \\
SU($F$)$^3$ & $Md_3(F)$ \\
SU($F$)$^2$U(1)$_R$ & $-M\frac{2M-N}{2F}d_2(F)$ \\   
SU($F$)$^2$U(1)$_B$ & $d_2(F)$\\   
SU($F$)$^2$U(1)$_X$ & 0 \\
SU($2G$)$^3$ & $Nd_3(2G)$   
\end{tabular}

\begin{tabular}{r l}      
SU($2G$)$^2$U(1)$_R$ & $-N\frac{N-M-2}{2G}d_2(2G)$ \\    
U(1)$_B^2$U(1)$_R$ & $-2$ \\   
U(1)$_X^2$U(1)$_R$ & $-MN$ \\   
U(1)$_R$U(1)$_B$U(1)$_X$ & $-N$    
\end{tabular}   
     
\vspace{0.5cm}   
   
\noindent where $d_3(F)$ and $d_2(F)$ are the cubic and quadratic $SU(F)$   
Casimirs of the fundamental representation. It is interesting to observe 
that the dual theory for $SU(M) \times SO(N)$ has much higher rank than  
the dual of an $SU(M) \times SU(N)$ theory with analogous matter 
content \cite{in}.

\section{Brane moves} \label{bm}   
\begin{figure}[hbt]   
\begin{center}   
\includegraphics[angle=0, width=0.8\textwidth]{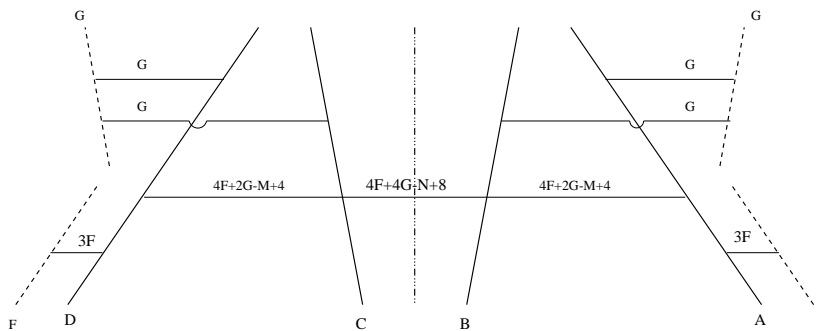}   
\caption{Brane configuration obtained by reversing the order of all branes.}   
\label{nauf2}   
\end{center}   
\end{figure}

\begin{figure}[hbt]   
\begin{center}   
\includegraphics[angle=0, width=0.8\textwidth]{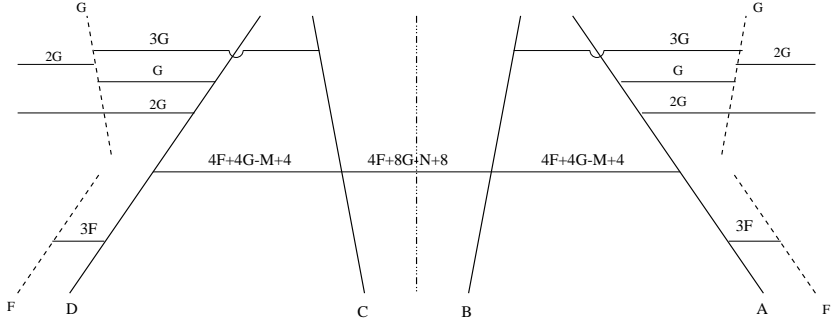}   
\caption{Dual brane configuration after adding $4G$ full fourbranes.}   
\label{nauf3}   
\end{center}   
\end{figure}

The electric brane configuration is shown in Fig.\ref{nauf1}. To find the   
dual theory, we reverse the order of the sixbranes as well as the   
fivebranes using   
the linking number conservation argument given in \cite{hw}. The orientifold   
plane is treated according to its Ramond charge, i.e. as a set of four  
sixbranes. With these rules we get the configuration shown in  
Fig.\ref{nauf2}.   
We obtain a gauge group $SU({\widetilde M}') \times SO({\widetilde N}')$ with   
$\widetilde{M}'=4F+2G-M+4$ and $\widetilde{N}'=4G+4F-N+8$, which does not  
coincide  
with that derived in the previous section. This mismatch can be   
cured by adding $4G$ full fourbranes to the dual configuration as shown in   
Fig.\ref{nauf3}.   
This does not affect the linking numbers.  
An analogous problem was encountered in \cite{hb} when rederiving the   
Seiberg dual for an $SU(N_1)\times SU(N_2) \times SU(N_3)$ gauge theory  
from brane moves. The brane configuration for that case is  
very similar to ours, it contains also four fivebranes. The number of  
fourbranes they had to add was twice the number of sixbranes placed  
between the second and the third fivebrane. We get the same result.   
  
In the rest of this section we want to propose an explanation for the  
necessity of adding full fourbranes to recover the conjectured dual 
theory\footnote{See   
also \cite{geometry} for a reinterpretation of the additional fourbranes   
using a different approach}. We will show that there exists a deformation  
of the electric theory that higgses the dual theory proposed in
section 3 down to the result derived from brane moves.   
Deformations of the electric theory superpotential associated to mesons  
correspond generically to higgsing in the dual theory. Particular cases of   
meson deformations are those generated by $M_0$ and $M'_0$, which give 
masses   
to the quarks $Q, \widetilde{Q}$ and $Q'$ of the electric theory. They have 
a simple geometrical interpretation which corresponds to change the position 
of the sixbranes in the orthogonal directions to the fivebranes. However 
deformations of the   
superpotential generated by higher mesons do not have a clear geometrical   
interpretation. Thus from the brane configuration used to derive the   
electric theory, we can not determine a priori if some   
of these deformations are switched on.   
  
In the dual brane configuration of Fig.\ref{nauf2} we have $3F$ fourbranes  
suspended between the A fivebrane and its set of parallel sixbranes.   
The fourbranes can slide in the two directions shared by the fivebrane and  
the sixbranes. This can be understood as giving expectation values to the   
diagonal components of the $M_i$ mesons, with $i=0,1,2$ \cite{elgvkt}.  
There are $G$ fourbranes connecting the B fivebrane and its set of parallel   
sixbranes. This number is sufficient to provide the $2G$ fields $q'$   
transforming in the vector representation of $SO(\widetilde{N})$, but it 
seems   
that some of the $M'_i$ mesons are missing. Based on this heuristic 
argument we will consider that the superpotential associated to the 
electric brane 
configuration of Fig.\ref{nauf1} is actually $W+ \Delta W$, with $W$ 
given by (\ref{w}) and $\Delta W$ a certain deformation generated by the 
mesons 
$M'_1$ and $M'_2$. The superpotential of the dual theory will thus be 
  \begin{eqnarray} \label{def}   
W &= & \mbox{Tr} (Y\widetilde{Y})^2+\mbox{Tr} Y\widetilde{Y}\widetilde{Y}Y  
+(\mbox{Tr}Y\widetilde{Y})^2+M'_1q'Y\widetilde{Y}q' \nonumber \\    
&& +M'_2q'q'- m_1 M'_1 - m_2 M'_2 + \dots ,  
  \end{eqnarray}   
where the dots stand for the other terms in (\ref{spdual}).  
  
The dual group $SU(\widetilde{M}) \times SO(\widetilde{N})$ differs from that  
obtained from brane moves by $\widetilde{M}=\widetilde{M}'+2G$,  
$\widetilde{N}=\widetilde{N}'+4G$. It will be sufficient to show that there   
exists a higgsing from $SU({\widetilde M}) \times SO({\widetilde N})$ to   
$SU({\widetilde M}-1) \times SO({\widetilde N}-2)$ without changing the 
matter   
content. Then by iterating this process $2G$ times we will arrive at the 
desired result. With this in mind we now study the superpotential 
(\ref{def}) when $(m_1)_{\alpha \beta}=  
(m_2)_{\alpha \beta}= \delta_{\alpha 1}  \delta_{\beta 1}$, where  
$\alpha$ and $\beta$ are $SU(2G)$ flavor indices. The F-term equations for   
$M'_1$ and $M'_2$ are  
\begin{equation}  
q'_{\alpha} Y {\widetilde Y} q'_{\beta} = \delta_{\alpha 1}  \delta_{\beta 1}  
\; , \hspace{3mm}  q'_{\alpha}q'_{\beta} = \delta_{\alpha 1}  
\delta_{\beta 1}.  
\label{fM}  
\end{equation}  
Assuming that all the singlet fields have zero expectation value, the F-term   
equations, (\ref{f1}) and (\ref{fM}), and the D-term equations are solved  
by  
\begin{eqnarray}  
\langle Y \rangle^a_i &=& \langle {\widetilde Y} \rangle_{a i} \;\:=\; \:  
\frac{1}{\sqrt 3} \: ( \: \delta_{a1} \delta_{i1} +i \delta_{a1} \delta_{i2}   
\:), \nonumber \\  
\langle q'_1 \rangle_i &=& \frac{1}{\sqrt 3} \:( \: 2\delta_{i1} -   
i \delta_{i2}\: ),  \label{ex}
\end{eqnarray}   
where $a=1,..,{\widetilde M}$ and $i=1,..,{\widetilde N}$ are the $SU$ and   
$SO$ indices respectively. These expectation values break the gauge group  
to $SU({\widetilde M}-1) \times SO({\widetilde N}-2)$ as expected. It 
remains to  
analyze the matter content of the higgsed theory.   
We get a bifundamental flavor for the higgsed theory from the fields $Y$,   
${\widetilde Y}$. The D-term equations allow also to recover from  $Y$,   
${\widetilde Y}$ an $SO({\widetilde N}-2)$ vector and an 
$SU({\widetilde M}-1)$ flavor in the fundamental representation. 
Substituting the expectation values (\ref{ex}) one can see that the 
superpotential gives mass to this additional $SU$   
flavor, but not to the $SO$ vector. Thus the higgsed theory has the  
same content of charged matter as the initial one, i.e. a bifundamental 
flavor,  
$F$ fundamental flavors of $SU({\widetilde M}-1)$ and $2G$ vector fields of  
$SO({\widetilde N}-2)$.  

We have considered a deformation generated by both $M'_1$ and 
$M'_2$. It is easy to see that setting $m_2=0$ does not alter the 
previous result, it only changes the particular expectation value of
$q'_1$. Thus it is a deformation generated by $M'_1$ that seems to 
be necessarily switched on in the brane construction. In order to
obtain a better understanding of this point, let us perform a brane move  
associated to dualize the $SU(M)$ factor group considering $SO(N)$ as a 
flavor group. 
\begin{figure}[hbt]   
\begin{center}   
\includegraphics[angle=0, width=0.8\textwidth]{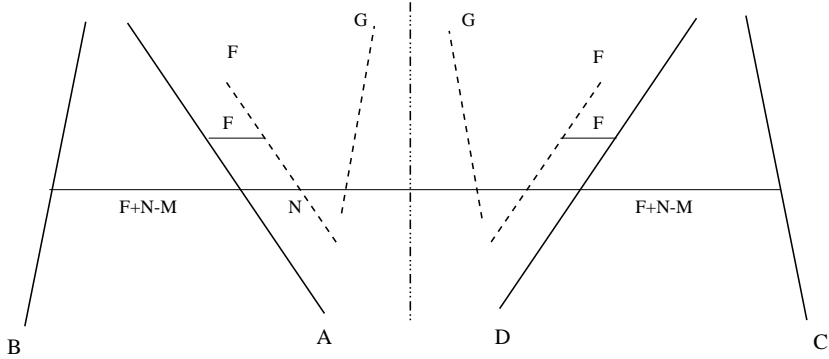}   
\caption{Brane configuration after interchanging A and B fivebranes.}   
\label{nauf4}   
\end{center}   
\end{figure}  
This is done by
moving the sixbranes parallel to the A fivebrane over B and then exchanging 
A and B as depicted in Fig.\ref{nauf4}. We have not moved the two groups 
of sixbranes in the central part of the diagram, which give rise to $2G$ 
$SO(N)$ vector fields, $Q'$. We observe that now the $2G$ central sixbranes
are not parallel to the fivebranes that define the $SO(N)$ group, i.e.
A and D. We perform now a further move
corresponding to dualize the $SO(N)$ factor group keeping $SU$ as 
a spectator.
This is done by bringing the sixbranes to the left of the orientifold
in Fig.\ref{nauf4} over the D fivebrane, the sixbranes to the right of the 
orientifold
over the A fivebrane and then exchanging A and D. The fourbranes 
created in this process between the two sets of $G$ sixbranes and the
A and D fivebranes can not slide along the world-volume of the fivebrane,
since the mentioned fivebranes and sixbranes are not parallel. In terms
of the dual $SO$ theory this means that the singlet field associated 
with the $SO(N)$ meson $Q'Q'$ must be massive. 
A mass term for this singlet can only be obtained when the superpotential 
of the 
electric $SO(N)$ theory contains a quartic term in the fields $Q'$.
An interesting remark is that Fig.\ref{nauf4} realizes only an $SU(G)$ 
subgroup of the $SU(2G)$ flavor symmetry for the $Q'$, and not 
$SU(G)\times SU(G)$ as Fig.\ref{nauf1} suggests.   

\begin{figure}[hbt]   
\begin{center}   
\includegraphics[angle=0, width=0.8\textwidth]{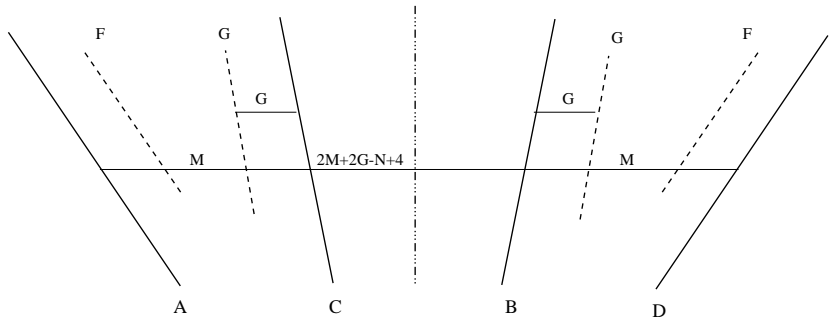}   
\caption{Brane configuration after interchanging B and C fivebranes.}   
\label{nauf5}   
\end{center}   
\end{figure}

Let us compare the previous situation with that obtained by
dualizing  instead the $SO(N)$ group in the configuration of 
Fig.\ref{nauf1}.
Then the fourbranes created 
between the two sets of $G$ sixbranes and the B and C fivebranes 
can slide along the world-volume of the fivebrane, since the central 
sixbranes have been assumed to be parallel to B and C (see 
Fig.\ref{nauf5}). In the dual
$SO$ theory there must be a massless singlet. Therefore a quartic 
term in $Q'$ will not be present in the electric theory\footnote{
There is a subtlety here. The mentioned singlet
transforms in the symmetric representation of $SU(2G)$, 
but only an $SU(G)$ subgroup of $SU(2G)$ is realized in the brane diagram. 
We can not rule out the presence of a restricted quartic term
in $Q'$ which would lift some components of the dual singlet 
reducing it to an $SU(G)$ adjoint.}.
However, according to the previous paragraph, the superpotential for 
the configuration in Fig.\ref{nauf1} should be such that when we dualize 
the $SU(M)$ factor group the resulting dual superpotential does indeed
contain a quartic term in the fields $Q'$. 

This is precisely achieved by adding to the superpotential (\ref{w})
a deformation generated by the meson $M'_1$
\begin{equation}
W= (X {\widetilde X})^2 + m_{\alpha \beta} \:
{Q'}^{\alpha} X {\widetilde X} {Q'}^{\beta},
\label{wm}
\end{equation}
where we have denoted all the terms appearing in (\ref{w}) as 
$(X {\widetilde X})^2$. 
Dualizing the $SU(M)$ factor group \cite{sei} we get an $SU(F+N-M) 
\times SO(N)$
gauge theory with a bifundamental flavor, $F$ $SU$ quark flavors and 
$2G+2F$ $SO$ vector fields. The additional $2F$ $SO$ vectors
have their origin in the $SU(M)$ chiral mesons $\widetilde{Q} X$ and $Q 
{\widetilde X}$. We will denote them by $q'_1$ and $q'_2$ respectively.
The $SO(N)$ fields $q'_1$ and $q'_2$ do have a representation 
in the brane diagram of Fig.\ref{nauf4}. Since the set of $F$ sixbranes 
is parallel to the A fivebrane, the fourbranes suspended between them 
can be moved arbitrarily far away from the intersection between the 
$F$ sixbranes and the $N$ central fourbranes. Thus strings joining the
$F$ sixbranes and the $N$ central fourbranes will give rise
to the $2F$ additional $SO$ vector fields. This conclusion is of course
not valid when the $F$ sixbranes and the A fivebrane are not parallel
\cite{hw}. The dual $SU(F+N-M) \times SO(N)$
theory contains also a singlet field $M_Q$ and a field $M_X$
transforming as the direct sum of the adjoint and the symmetric 
representations of $SO(N)$. These fields are in correspondence with the 
$SU(M)$ chiral mesons $Q\widetilde{Q}$ and $X {\widetilde X}$. The quartic 
superpotential for the bifundamental fields in (\ref{wm}) translates into
a mass term for $M_X$ in the dual theory
\begin{equation}
W= (M_X)^2 + m\: Q' M_X Q' + M_Q q \tilde{q} + q'_1 Y \tilde{q} + q'_2
\: \widetilde{Y} q +  M_X Y{\widetilde Y}.
\end{equation}  
Integrating out this field, we get
\begin{equation}
W= (Y \widetilde{Y} + m \: Q' Q')^2 +  M_Q q \tilde{q} + q'_1 Y \tilde{q} 
+ q'_2 \widetilde{Y} q,
\label{qq}
\end{equation}
which contains a quartic term for the fields $Q'$.

We can obtain more information about the matrix $m_{\alpha 
\beta}$ by considering again the situation 
of Fig.\ref{nauf5}, which corresponds to dualize the $SO(N)$ factor 
group. The resulting dual group is $SU(M) \times SO(2M+2G-N+4)$
\cite{sei} \cite{sint}. The brane diagram implies that there are 
$F+G$ $SU(M)$ quarks. The additional $SU$ quarks can only have their 
origin in the $SO(N)$ chiral mesons $Q'X$, $Q' \widetilde{X}$. 
We denote them by $q$ and $\tilde q$ respectively; they form $2G$
$SU(M)$ flavors. The term $m Q' X \widetilde{X} Q'$ in (\ref{wm})
translates in the dual $SU(M) \times SO(2M+2G-N+4)$ theory into a
mass term for $q$ and $\tilde q$
\begin{equation}
m_{\alpha \beta} q^{\alpha} {\tilde q}^{\beta}.
\end{equation}
Thus $m$ has to be such that it gives mass to $G$ of the new
flavors. A way of choosing $m$ with the above property and
preserving an $SU(G)$ subgroup of 
the $SU(2G)$ flavor symmetry is
\begin{equation}
m_{\alpha \beta} \sim \left\{ \begin{array}{ll}
\delta_{\alpha +G , \beta} & \alpha=1,..,G, \\
0 & \textrm{otherwise}.
\end{array} \right.
\end{equation}
Notice that this expression for $m$ makes sense because the
meson $M'_1$ transforms as the direct sum of the symmetric and
antisymmetric representations of $SU(2G)$. 
A check for the proposed structure of $m$ is the following. When 
substituted in (\ref{qq}) and after dualizing further the $SO(N)$ group, 
it must induce 
mass terms for all the components of the singlet associated with
the $SO(N)$ meson $Q'Q'$. Remembering that the term in
parenthesis in (\ref{qq}) was a short way of denoting 
several terms as in (\ref{w}), we obtain
\begin{equation}
\begin{array}{lll}
m_{\alpha \beta} {Q'}_i^{\alpha}   {Q'}_j^{\beta}  m_{\delta \gamma}
{Q'}_j^{\delta}  {Q'}_i^{\gamma} & \sim & M^{\hat{\alpha},\hat{\delta}+G}  
M^{\hat{\delta},\hat{\alpha}+G},  \\
m_{\alpha \beta} {Q'}_i^{\alpha}   {Q'}_j^{\beta}  m_{\gamma \delta}
{Q'}_j^{\delta}  {Q'}_i^{\gamma} & \sim & M^{\hat{\alpha},\hat{\delta}}  
M^{\hat{\alpha}+G,\hat{\delta}+G}.
\end{array}
\end{equation}
with $\hat{\alpha} , \hat{\delta} =1,..,G$, which indeed gives mass 
to all components of $M$.

The previous arguments can be applied to more generic brane 
configurations than those considered here. They suggest that, for
configurations with several fivebranes and sixbranes, 
the superpotential includes generically quartic couplings between
quarks and bifundamental fields. The presence of such terms 
does not sound surprising for brane configurations with sixbranes
not parallel to the adjacent fivebranes. Our main result is that
for configurations in which the sixbranes are parallel to one of 
the adjacent fivebranes, the superpotential also contains a term $Q 
\widetilde{X} X \widetilde{Q}$ coupling the quarks coming from the 
sixbranes and the bifundamental field coming from the 
parallel fivebrane. In particular for configurations with more
than three fivebranes, such terms are unavoidable.
These terms translate in the dual theory into terms linear in the 
singlet fields, which have the effect of higgsing. 
Besides the case treated in this paper, this explains why 
the brane approach to $SU(N_1) \times SU(N_2) \times SU(N_3)$ 
predicts a dual group of smaller rank than that derived by field 
theory arguments \cite{hb}. 

We would like to end this section with one additional comment.
We have argued that the brane construction of our $SU(M) \times 
SO(N)$ theory corresponds to the modified superpotential (\ref{wm}) 
instead of (\ref{w}). However the brane 
moves necessary to derive the dual theory have a field theory 
interpretation independent of what the concrete superpotential is. 
They can be seen as successive, seperate dualizations of the $SU$ and $SO$ 
gauge groups. In particular the brane moves that bring us 
from Fig.\ref{nauf1} to Fig.\ref{nauf2} correspond to dualize first 
$SU$, then $SO$, then again $SU$ and finally again $SO$ (or
alternatively first $SO$, then $SU$ and then again $SO$ and $SU$).
We can apply this chain of dualities to the $SU(M) \times 
SO(N)$ theory with the undeformed superpotential (\ref{w}).
After each step fields transforming in tensor representations appear.
They are massive due to the quartic term in the bifundamental
fields in the superpotential and can be integrated out. Thus
we only need to use the known dualities for $SU$ and $SO$ groups
with matter in the fundamental and vector representation 
respectively \cite{sei}, \cite{sint}.
We have checked that the dual theory derived in this way 
coincides with the one proposed in section 3, which is a very strong
test for our conjectured dual theory\footnote{We thank the referee 
for suggesting this test to us.}. These calculations are 
straightforward but rather lengthy and we will not include them here.
However the first step, 
corresponding to dualize the $SU(M)$ group, has been explicitly
analyzed above with the modified superpotential.
  
\section{$SU(M) \times Sp(2N)$}  
  
We state briefly some results for the brane set-up in Fig. \ref{nauf1} 
with an  
orientifold sixplane of negative Ramond charge. In this case we obtain an 
$\mathcal{N}$=1 theory with gauge group $SU(M) \times Sp(2N)$ and the same 
matter content as before. The superpotential derived from the brane 
configuration is  
\begin{equation}  
W = a \mbox{Tr}( X {\widetilde X} )^2 + b  
\mbox{Tr} X {\widetilde X} {\widetilde X} X + c (\mbox{Tr} X 
{\widetilde X} )^2,  
\label{wsp}  
\end{equation}   
where  
\begin{equation}  
a=-\frac{1}{4} \! \left( \! \frac{1}{\mbox{tan} (\theta_2 \! - \! 
\theta_1)} +   
\frac{1}{\mbox{tan} 2 \theta_1} \! \right),  
\hspace{3mm} b=\frac{-1}{4 \mbox{sin} 2 \theta_1}, \hspace{3mm}  
c=\frac{1}{4 M \mbox{tan} (\theta_2 \! - \! \theta_1)}.  
\label{valuessp}  
\end{equation}  
The mesons are the ones given in section \ref{bc} and also the global 
symmetry group. The transformation properties of the matter fields under 
the gauge and global symmetry groups are listed in table \ref{tspe}.  
\begin{table}[hbt!]   
\begin{center} 
\begin{small}  
\setlength{\tabcolsep}{1mm}  
\begin{tabular}{|c||c|c|c|c|c|c|c|c|}   
\hline    
 & SU($M$) & Sp($2N$) & SU($F$)$_L$ & SU($F$)$_R$ & SU($2G$) & U(1)$_R$ 
& U(1)$_B$ & U(1)$_X$ \\   
\hline \hline   
$Q$ & $M$ & 1 & $F$ & 1 & 1 & $1-\frac{M-N}{F}$ & $\frac{1}{M}$ & 0  \\   
\hline   
$\widetilde{Q}$ & $\overline{M}$ & 1 & 1 & $F$ & 1 & $1-\frac{M-N}{F}$ 
& -$\frac{1}{M}$ & 0 \\   
\hline   
$Q'$ & 1 & $2N$ & 1 & 1 & $2G$ & $1-\frac{2N-M+2}{2G}$ & 0 & 0 \\   
\hline   
$X$ & $M$ & $2N$ & 1 & 1 & 1 & $\frac{1}{2}$ & $\frac{1}{M}$ & 1 \\   
\hline   
$\widetilde{X}$ & $\overline{M}$ & $2N$ & 1 & 1 & 1 & $\frac{1}{2}$ & 
-$\frac{1}{M}$ & -1 \\   
\hline   
\end{tabular} 
\end{small}   
\caption{Matter content of the electric theory.} \label{tspe}    
\end{center}   
\end{table}   

The dual theory has gauge group $SU(\widetilde{M})\times 
Sp(2\widetilde{N})$ with $\widetilde{M}=4F+4G-M-4$ and 
$\widetilde{N}=2F+4G-N-4$. The field content of the dual theory and
the transformation under the symmetries are indicated in table \ref{tspm}.   
\begin{table}[hbt!]   
\begin{center} 
\begin{small}   
\setlength{\tabcolsep}{1mm}  
\begin{tabular}{|c||c|c|c|c|c|c|c|c|}     
\hline    
 & SU(${\widetilde M}$) & Sp(${2\widetilde{N}}$) & SU($F$)$_L$ & 
SU($F$)$_R$ & SU($2G$) & U(1)$_R$ & U(1)$_B$ & U(1)$_X$ \\   
\hline \hline   
$q$ & $\widetilde{M}$ & 1 & $\overline{F}$ & 1 & 1 & 
$1-\frac{\widetilde{M}-\widetilde{N}}{F}$ & 
$\frac{1}{\widetilde{M}}$ &$\frac{\widetilde{M}+M-2F}{\widetilde{M}}$  \\   
\hline   
$\widetilde{q}$ & $\overline{\widetilde{M}}$ & 1 & 1 & 
$\overline{F}$ & 1 & $1-\frac{\widetilde{M}-\widetilde{N}}{F}$ & 
-$\frac{1}{\widetilde{M}}$ & 
$-\frac{\widetilde{M}+M-2F}{\widetilde{M}}$  \\   
\hline   
$q'$ & 1 & $2\widetilde{N}$ & 1 & 1 & $\overline{2G}$ & 
$1-\frac{2\widetilde{N}-\widetilde{M}+2}{2G}$ & 0 & 0 \\   
\hline   
$Y$ & $\widetilde{M}$ & $2\widetilde{N}$ & 1 & 1 & 1 & $\frac{1}{2}$ & 
$\frac{1}{M}$ & $\frac{M-2F}{\widetilde{M}}$ \\   
\hline   
$\widetilde{Y}$ & $\overline{\widetilde{M}}$ & $2\widetilde{N}$ & 1 & 1 
& 1 & $\frac{1}{2}$ & -$\frac{1}{\widetilde{M}}$ & 
$-\frac{M-2F}{\widetilde{M}}$ \\   
\hline   
\end{tabular} 
\end{small}   
\caption{Matter content of the magnetic theory.} \label{tspm}    
\end{center}   
\end{table}   
Note that the mesons $M'_0$ and $M'_2$ are now in the antisymmetric  
representation of $SU(2G)$. The dual theory has a superpotential as in  
(\ref{spdual}). As in the previous case the dual theory can be obtained
from the known dualities for $SU$ and $Sp$ groups \cite{sei}, \cite{inp}
by dualizing first the $SU$ factor, then the $Sp$ factor and then $SU$ 
and $Sp$ again\footnote{ 
The dualities for $SU$ and $Sp$ with matter in the fundamental 
representation are derived in \cite{sei}, \cite{inp} for zero 
superpotential. We have however a non-zero superpotential which
induces different global symmetries from those of the $W\!=\!0$ case. 
Thus we have explicitly checked the t'Hooft anomaly matching conditions,
which are indeed satisfied.} (or
alternatively first $Sp$, then $SU$ and then again $Sp$ and $SU$).     
 
When we try to recover the dual theory from brane moves we get a smaller 
dual group $SU(\widetilde{M}') \times Sp(2\widetilde{N}')$ 
with $\widetilde{M}'=\widetilde{M}-2G$, $\widetilde{N}'=\widetilde{N}-2G$. 
We can cure this mismatch by adding $4G$ full fourbranes to the dual 
configuration. All the arguments presented in the previous section 
to explain this problem extend to the $SU \times Sp$ case.
We can understand the addition of the fourbranes as a reverse of higgsing 
in the dual theory, induced by adding to the superpotential (\ref{wsp})
of the electric theory a deformation generated by the meson $M'_1$.

\begin{center} {\bf Acknowledgments} \end{center}    
We thank K. Landsteiner for discussions. The work of E.L. is
supported by a Lise Meitner Fellowship, M456-TPH. The
work of B.O. is supported by the \"Osterreichische
Nationalbank
Jubil\"aumsfonds under contract 6584/4.

\end{document}